\documentclass[12pt]{article}
\usepackage{mathrsfs} \usepackage{amsmath} \usepackage{amssymb} \usepackage{slashed}
\usepackage{graphicx} \usepackage{caption2}
\begin{document}
\newcommand{\ba}{\begin{eqnarray}} \newcommand{\ea}{\end{eqnarray}}
\newcommand{\be}{\begin{equation}} \newcommand{\ee}{\end{equation}}
\renewcommand{\figurename}{Figure.} \renewcommand{\captionlabeldelim}{.~}
\renewcommand{\thefootnote}{\fnsymbol{footnote}}

\vspace*{1cm}
\begin{center}
 {\Large\textbf{A Model of Dark Matter, Leptogenesis, and Neutrino Mass from the $B-L$ Violation just above the Electroweak Scale}}

 \vspace{1cm}
 \textbf{Wei-Min Yang}

 \vspace{0.3cm}
 \emph{Department of Modern Physics, University of Science and Technology of China, Hefei 230026, P. R. China}

 \emph{E-mail: wmyang@ustc.edu.cn}
\end{center}

\vspace{1cm}
\noindent\textbf{Abstract}: I suggest an extension of the SM by introducing a dark sector with the local $U(1)_{D}$ symmetry. The particles in the dark sector bring about the new physics beyond the SM. In particular the global $B-L$ symmetry is violated just above the electroweak scale, this becomes a common origin of the tiny neutrino mass, the cold dark mater and the baryon asymmetry. The model can not only account for the tiny neutrino mass and the ``WIMP Miracle", but also achieve the leptogenesis around the electroweak scale. Finally, it is very possible that the model predictions are tested in near future experiments.

\vspace{1cm}

\newpage
\noindent\textbf{I. Introduction}

\vspace{0.3cm}
  The standard model (SM) of the particle physics has successfully accounted for all kinds of the physics at or below the electroweak scale, refer to the reviews in Particle Data Group \cite{1}, but it can not explain the three important issues: the tiny neutrino mass \cite{2}, the cold dark matter (CDM) \cite{3}, and the matter-antimatter asymmetry \cite{4}. Many theories have been suggested to solve these problems. The tiny neutrino mass can be generated by the seesaw mechanism \cite{5} or the other means \cite{6}. The baryon asymmetry can be achieved by the thermal leptogenesis \cite{7} or the electroweak baryogenesis \cite{8}. The CDM candidates are possibly the sterile neutrino \cite{9}, the lightest supersymmetric particle \cite{10}, the axion \cite{11}, and so on. In addition, some inspired ideas attempt to find some connections among the neutrino mass, the CDM, and the baryon asymmetry, for example, the lepton violation can lead to the neutrino mass and the baryon asymmetry \cite{12}, the neutrino mass and the leptogenesis are implemented by the heavy scalar triplet \cite{13}, the asymmetric CDM is related to the baryon asymmetry \cite{14}, and some models unifying them into a frame \cite{15}. Although many progresses on these fields have been made all the time, an universal and convincing theory is not established as yet.

  The universe harmony and the nature unification are a common belief of mankind. It is hard to believe that the tiny neutrino mass, the CDM and the matter-antimatter asymmetry appear to be not related to each other, conversely, it is very possible that the three things have a common origin. Therefore, a new theory beyond the SM should be capable of accounting for the three things simultaneously. On the other hand, a realistic theory should keep such principles as the simplicity and the fewer number of parameters, in addition, it should be feasible and promising to be tested in future experiments. If one theory is excessive complexity and unable to be tested, it is unbelievable and infeasible. Based on these considerations, I suggest a new extension of the SM. It only introduces a few of new particles with a local gauge symmetry of $U(1)_{D}$, which are in the dark sector. In particular, the global symmetry of $U(1)_{B-L}$ is violated just above the electroweak scale, this becomes a common origin of the above three things. The model can simply and completely account for the above three issues, and it is very feasible to test the model by the TeV-scale colliders, the underground detectors, and the search in the cosmic rays.

  The remainder of this paper is organized as follows. I outline the model In Section II. Section III and Section IV are respectively discussions of the dark matter and the leptogenesis. I give the numerical results and discuss the model test in Section V. Section VI is devoted to conclusions.

\vspace{0.6cm}
\noindent\textbf{II. Model}

\vspace{0.3cm}
  The model introduces a local gauge symmetry $U(1)_{D}$ and some new particles with the $D$ numbers besides the SM sector, in addition, it keeps the global symmetry $U(1)_{B-L}$, i.e., the difference of the baryon number and the lepton one is conserved. The model particle contents and their gauge quantum numbers under $SU(2)_{L}\otimes U(1)_{Y}\otimes U(1)_{D}$ are listed as follows,
\begin{alignat}{1}
 &l_{L}(2,-1,0)_{-1},\hspace{0.3cm} e_{R}(1,-2,0)_{-1},\hspace{0.3cm} N_{R}(1,0,0)_{-1},\hspace{0.3cm} N_{L}(1,0,1)_{-1},\hspace{0.3cm} \chi_{R}(1,0,1)_{-1}, \nonumber\\
 &H(2,-1,0)_{0}\,,\hspace{0.3cm} \Phi(2,-1,1)_{-2}\,,\hspace{0.3cm} \phi_{1}(1,0,1)_{0}\,,\hspace{0.3cm} \phi_{2}(1,0,2)_{-2}\,,
\end{alignat}
where the right subscripts of the brackets indicate the $B-L$ numbers under $U(1)_{B-L}$. Here I omit the quark sector and the color subgroup $SU(3)_{C}$ since what followed have nothing to do with them. The particles in the SM sector have no any $D$ numbers, while those particles with the non-vanishing $D$ numbers are in the dark sector. Note that $N_{R}$ is filled into the SM sector but $N_{L}$ belongs to the dark sector. All of the fermions in Eq. (1) have three generations as usual. It is easily verified that all the chiral anomalies are completely cancelled by virtue of the assignment of Eq. (1), namely the model is anomaly-free. We also understand the model symmetries from another point. We can infer that $U(1)_{Y}$ is essentially derived from a linear combination of $U(1)_{D}$, $U(1)_{B-L}$, and a hidden gauge symmetry $U(1)_{I^{R}_{3}}$. The relation of their quantum numbers is
\ba
\frac{Y}{2}=I^{R}_{3}+\frac{B-L}{2}+\frac{D}{2}\,.
\ea
The assignment of $I^{R}_{3}$ is as follows, $I^{R}_{3}=\frac{1}{2}$ for $N_{R}$, $I^{R}_{3}=-\frac{1}{2}$ for $e_{R},H, \phi_{1}$, and $I^{R}_{3}=0$ for the other fields of Eq. (1). Thus $U(1)_{Y}$ is regarded as a relic of the breakings of the above three Abelian subgroups. Finally, the model has also a hidden $Z_{2}$ symmetry, it is defined by the following transform
\ba
f_{L}\rightarrow-f_{L},\hspace{0.3cm} f_{R}\rightarrow f_{R},\hspace{0.3cm} H\rightarrow-H,\hspace{0.3cm} \Phi\rightarrow\Phi,\hspace{0.3cm} \phi_{1}\rightarrow-\phi_{1},\hspace{0.3cm} \phi_{2}\rightarrow\phi_{2},
\ea
where $f_{L,R}$ denote the left-handed and right-handed fermions in Eq. (1). Note that $N_{L}$ and $\chi_{R}$ have the same gauge quantum numbers but they have opposite $Z_{2}$ parities.

  Under the above symmetries, the invariant Lagrangian of the model is composed of the three following parts. The gauge kinetic energy terms are
\begin{alignat}{1}
 \mathscr{L}_{G}&=\mathscr{L}_{pure\:gauge}+\sum\limits_{f}i\,\overline{f}\,\gamma^{\mu}D_{\mu}f+\sum\limits_{S}(D^{\mu}S)^{\dagger}D_{\mu}S\,, \nonumber\\
 D_{\mu}&=\partial_{\mu}+i\left(g_{2}W_{\mu}^{i}\frac{\tau^{i}}{2}+g_{1}B_{\mu}\frac{Y}{2}+g_{0}X_{\mu}\frac{D}{2}\right),
\end{alignat}
where $f$ and $S$ respectively denote all kinds of the fermions and scalars in Eq. (1), $g_{0}$ and $X_{\mu}$ is the gauge coupling coefficient and gauge field associated with $U(1)_{D}$, the other notations are self-explanatory.

  The Yukawa couplings are
\begin{alignat}{1}
 \mathscr{L}_{Y}=&\:\overline{l_{L}}Y_{e}e_{R}\,i\tau_{2}H^{*}+\overline{l_{L}}Y_{1}N_{R}H+\overline{l_{L}}Y_{2}C\overline{N_{L}}^{T}\Phi \nonumber\\
 &+\overline{N_{L}}Y_{N}N_{R}\,\phi_{1}+\frac{1}{2}N_{L}^{T}CY'_{N}N_{L}\phi^{*}_{2}+\frac{1}{2}\chi_{R}^{T}CY_{\chi}\chi_{R}\,\phi^{*}_{2}+h.c.\,,
\end{alignat}
where $\tau_{2}$ is the second Paul matrix and $C$ is the charge conjugation matrix. Note that the $Z_{2}$ symmetry of Eq. (3) forbids the explicit mass term $\overline{N_{L}}M\chi_{R}$ even though it satisfies all the gauge symmetries. The coupling parameters have reasonable size as $[Y_{e},Y_{1},Y_{2}]\sim0.01$ and $[Y_{N},Y'_{N},Y_{\chi}]\sim0.1$. They are generally $3\times3$ complex matrices in the flavour space, however, we can choose such flavour basis in which $Y_{e},Y_{N},Y_{\chi}$ are simultaneously diagonal matrices (namely the mass eigenstate basis, see Eq. (9) below), thus $Y_{1}$ and $Y_{2}$ certainly contain some irremovable complex phases, they eventually become $CP$-violating sources in the lepton sector in comparison with one in the quark sector. Eq. (5) will give rise to all kinds of the fermion masses after the scalar fields developing their non-vanishing vacuum expectation values. After the sequential breakings of $U(1)_{D}$ and $U(1)_{B-L}$, the $Y_{1}$ and $Y_{2}$ terms will lead to the tiny neutrino mass and the leptogenesis, and the $Y_{\chi}$ term will generate the CDM.

  The full scalar potentials are
\begin{alignat}{1}
 V_{S}=&\:\frac{1}{4\lambda_{\Phi}}\left(2\lambda_{\Phi}\Phi^{\dagger}\Phi-\left(\lambda_{\Phi}v_{\Phi}^{2}+(\lambda_{0}+\lambda_{1})v_{H}^{2}+\lambda_{2}v_{1}^{2}+\lambda_{3}v_{2}^{2}\right)+(\frac{M_{\Phi}v_{\Phi}}{v_{\Phi}})^{2}\right)^{2} \nonumber\\
 &+\frac{1}{4\lambda_{H}}\left(2\lambda_{H}H^{\dagger}H-\left((\lambda_{0}+\lambda_{1})v_{\Phi}^{2}+\lambda_{H}v_{H}^{2}+\lambda_{4}v_{1}^{2}+\lambda_{5}v_{2}^{2}\right)+(\frac{M_{\Phi}v_{\Phi}}{v_{H}})^{2}\right)^{2} \nonumber\\
 &+\frac{1}{4\lambda_{\phi_{1}}}\left(2\lambda_{\phi_{1}}\phi^{*}_{1}\phi_{1}-\left(\lambda_{2}v_{\Phi}^{2}+\lambda_{4}v_{H}^{2}+\lambda_{\phi_{1}}v_{1}^{2}+\lambda_{6}v_{2}^{2}\right)+(\frac{M_{\Phi}v_{\Phi}}{v_{1}})^{2}\right)^{2} \nonumber\\
 &+\frac{1}{4\lambda_{\phi_{2}}}\left(2\lambda_{\phi_{2}}\phi^{*}_{2}\phi_{2}-\left(\lambda_{3}v_{\Phi}^{2}+\lambda_{5}v_{H}^{2}+\lambda_{6}v_{1}^{2}+\lambda_{\phi_{2}}v_{2}^{2}\right)+(\frac{M_{\Phi}v_{\Phi}}{v_{2}})^{2}\right)^{2} \nonumber\\
 &+2\lambda_{0}\Phi^{\dagger}HH^{\dagger}\Phi+2\Phi^{\dagger}\Phi\left(\lambda_{1}H^{\dagger}H+\lambda_{2}\phi^{*}_{1}\phi_{1}+\lambda_{3}\phi^{*}_{2}\phi_{2}\right) \nonumber\\
 &+2H^{\dagger}H\left(\lambda_{4}\phi^{*}_{1}\phi_{1}+\lambda_{5}\phi^{*}_{2}\phi_{2}\right)+2\lambda_{6}\phi^{*}_{1}\phi_{1}\phi^{*}_{2}\phi_{2}-2\left(\lambda_{7}\Phi^{\dagger}H\phi^{*}_{1}\phi_{2}+h.c.\right),
\end{alignat}
where $v_{\Phi}=\frac{\lambda_{7}v_{H}v_{1}v_{2}}{M_{\Phi}^{2}}$. Note that $v_{\Phi}$ is not an independent parameter in Eq. (6), in fact, there are only four independent mass-dimensional parameters, namely $[M_{\Phi},v_{H},v_{1},v_{2}]>0$, in which $M_{\Phi}$ is the original masses of $\Phi$ and the others are the vacuum expectation values (see Eq. (8) below). These mass-dimensional parameters are assumed to be a hierarchy as
\ba
v_{\Phi}\sim 1\:\mathrm{MeV}\ll v_{H}\sim v_{2}\sim 300\:\mathrm{GeV}<M_{\Phi}\sim 5\:\mathrm{TeV}\ll v_{1}\sim 5000\:\mathrm{TeV}.
\ea
Those self-coupling parameters in Eq. (6) satisfy such conditions as $[\lambda_{\Phi},\lambda_{H},\lambda_{\phi_{1}},\lambda_{\phi_{2}}]\sim0.1>0$, while those interactive coupling parameters are assumed as $[\lambda_{0},\lambda_{1},\ldots,\lambda_{7}]\ll1$, for instance, $\lambda_{7}\sim10^{-7}$ is required by Eq. (7). In a word, the self-interaction of each scalar is strong but the interactions among them are very weak. However, the above conditions are natural and reasonable, they can sufficiently guarantee the vacuum stability. From a mathematical discussion of the minimum of $V_{S}$, we can rigorously derive the vacuum configurations as follows,
\ba
 \langle\Phi\rangle=\frac{v_{\Phi}}{\sqrt{2}}\left(\begin{array}{c}1\\0\end{array}\right),\hspace{0.3cm} \langle H\rangle=\frac{v_{H}}{\sqrt{2}}\left(\begin{array}{c}1\\0\end{array}\right),\hspace{0.3cm} \langle\phi_{1}\rangle=\frac{v_{1}}{\sqrt{2}}\,,\hspace{0.3cm} \langle\phi_{2}\rangle=\frac{v_{2}}{\sqrt{2}}\,.
\ea
$v_{H}=246$ GeV has been fixed by the electroweak physics. $v_{2}$ will be determined by the CDM. $v_{1}$ and $v_{\Phi}$ (or $v_{1}$ and $M_{\Phi}$) will be jointly determined by the tiny neutrino mass and the leptogenesis.

  Eq. (7) indicates the sequence of the symmetry breakings. Firstly $\langle\phi_{1}\rangle$ breaks the local $U(1)_{D}$ and the discrete $Z_{2}$, the neutral $N$ becomes a Dirac fermion with a mass around the $v_{1}$ scale. Secondly $\langle\phi_{2}\rangle$ violates the global $U(1)_{B-L}$, the neutral $\chi$ becomes a Majorana fermion with a mass around the $v_{2}$ scale. Thirdly the $SU(2)_{L}\otimes U(1)_{Y}$ breaking is accomplished by $\langle H\rangle$, the SM fermions obtain their masses around the electroweak scale. Note that the $B-L$ violation is just before the electroweak breaking due to $v_{2}\sim v_{H}$. Lastly $\Phi$ is induced developing a relatively small $\langle\Phi\rangle$ by the above three breakings, thus the tiny neutrino mass is generated by the seesaw mechanism after the heavy Dirac fermion $N$ is integrated out. All kinds of the particles masses are given as follows,
\begin{alignat}{1}
 &M_{X_{\mu}}=\frac{v_{1}g_{0}}{2}\,,\hspace{0.3cm} M_{\phi^{0}_{1}}=v_{1}\sqrt{2\lambda_{\phi_{1}}}\,,\hspace{0.3cm} M_{\phi^{0}_{2}}=v_{2}\sqrt{2\lambda_{\phi_{2}}}\,,\hspace{0.3cm} M_{G^{0}}=0,\hspace{0.3cm} M_{H^{0}}=v_{H}\sqrt{2\lambda_{H}}\,, \nonumber\\
 &M_{N}=-\frac{v_{1}}{\sqrt{2}}Y_{N},\hspace{0.5cm} M_{\chi}=-\frac{v_{2}}{\sqrt{2}}Y_{\chi},\hspace{0.5cm} M_{e}=\frac{v_{H}}{\sqrt{2}}Y_{e}, \nonumber\\
 &M_{\nu}=-\frac{v_{H}v_{\Phi}}{2}Y_{1}M_{N}^{-1}Y_{2}^{T}=\frac{v_{H}^{2}v_{2}}{\sqrt{2}M_{\Phi}^{2}}\lambda_{7}Y_{1}Y_{N}^{-1}Y_{2}^{T}.
\end{alignat}
Note that the real and imaginary parts of $\phi_{2}$, respectively, now become the massive neutral scalar boson denoted by $\phi^{0}_{2}$ and the massless Goldstone boson denoted by $G^{0}$. The mixing angle between $H^{0}$ and $\phi^{0}_{2}$ is $tan2\theta=\frac{2\lambda_{5}\,v_{H}v_{2}}{\lambda_{H}\,v_{H}^{2}-\lambda_{\phi_{2}}v_{2}^{2}}\ll 1$ due to $\lambda_{5}\ll 1$. For the weak couplings between the scalar bosons, all the mixings among them are very small and can be neglected. The mixing between $Z_{\mu}$ and $X_{\mu}$ is nearly zero since $\frac{v_{\Phi}^{2}}{v_{1}^{2}}$ is too small. Therefore, we can leave out all the mixings in the boson sector except the SM weak gauge mixing. In the fermion sector, the neutrino mass matrix $M_{\nu}$ bears all information of the neutrino mass and the lepton mixing.

  Based on Eq. (7) and Eq. (9), and we take into account of the mass hierarchy of $N_{1,2,3}$ and one of $\chi_{1,2,3}$, a reasonable mass spectrum relation for the model particles is such as (GeV as unit),
\begin{alignat}{1}
 &M_{G^{0}}<M_{\nu}\sim 10^{-10}\ll M_{e}<M_{\chi_{1}}\sim 10<M_{\chi_{2}}<M_{\chi_{3}}\sim M_{H^{0}}\sim M_{\phi^{0}_{2}}\sim 10^{2} \nonumber\\
 &<M_{\Phi}\sim 10^{3}<M_{N_{1}}\sim 10^{5}<M_{N_{2}}<M_{N_{3}}\sim M_{\phi^{0}_{1}}\sim M_{X_{\mu}}\sim 10^{6}.
\end{alignat}
This is easily satisfied by choosing some suitable values of the coupling parameters in Eq. (9). The mass relations of Eq. (10) will successfully lead to the CDM and the leptogenesis. Finally, it should be stressed that there are no any super-high scale physics in the model.

\vspace{0.6cm}
\noindent\textbf{III. Dark Matter}

\vspace{0.3cm}
  In the model, $\chi_{1,2,3}$ have no any interactions with the SM sector due to the $U(1)_{D}$ symmetry, in addition, they can not mix with $N_{1,2,3}$ due to the $Z_{2}$ symmetry, these features guarantee they are stable particles without any decays. After the $B-L$ symmetry is broken, $\chi_{1,2,3}$ justly become WIMPs. In the early universe, $\chi_{1,2,3}$ are in thermal equilibrium with the other particles in the dark sector. Afterwards the heavier $\chi_{2,3}$ mainly annihilate into the lightest $\chi_{1}$ via the $G^{0}$ mediator, shown as (a) in Fig. 1, eventually, $\chi_{1}$ annihilates into $G^{0}$ by the two modes of (b) and (c) in Fig. 1. After some careful analysis, the annihilation cross-sections of $\chi_{2,3}$ are much larger than one of $\chi_{1}$, in addition, $\chi_{2,3}$ have almost been decoupling before the $\chi_{1}$ annihilations take place. Therefore, the relic abundances of $\chi_{2,3}$ are much smaller than one of $\chi_{1}$, in other words, $\chi_{1}$ should be the principal particle of the CDM, while $\chi_{2,3}$ only bear a tiny part of the CDM budget. In short, $\chi_{1}$ is a desirable candidate of the CDM because its natures and relic abundance are very well consistent with ones of the CDM.

  After $\chi_{1}$ becomes non-relativistic particle, it has two annihilation channels, (i) $\chi_{1}+\overline{\chi_{1}}\rightarrow G^{0}+G^{0}$ via the t-channel mediation of $\chi_{1}$, shown as (b) in Fig. 1, (ii) $\chi_{1}+\chi_{1}\rightarrow G^{0}+G^{0}$ via the s-channel mediation of $\phi_{2}^{0}$, shown as (c) in Fig. 1.
\begin{figure}
 \centering
 \includegraphics[totalheight=8cm]{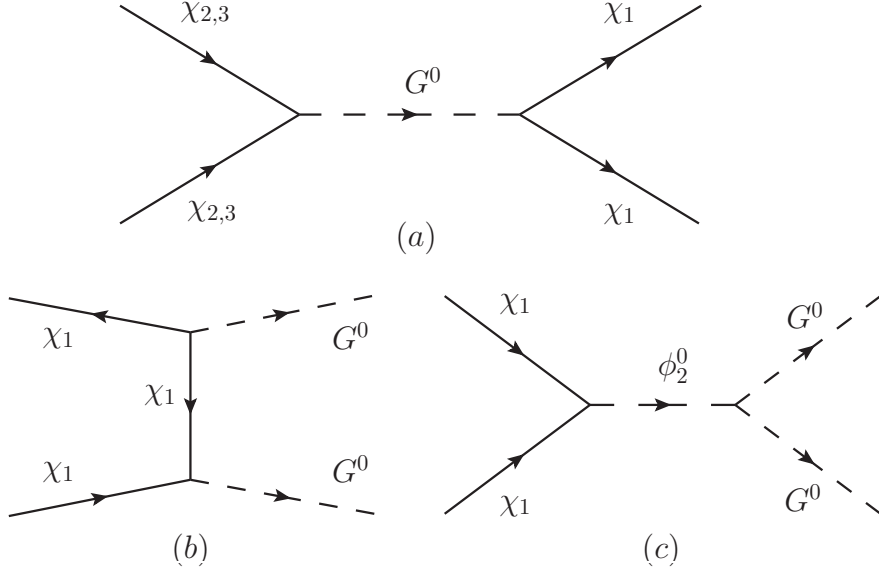}
 \caption{(a) A pair of heavier $\chi_{2,3}$ annihilating into a pair of the lightest $\chi_{1}$, (b) and (c) the pair annihilation of the CDM $\chi_{1}$ into a pair of Goldstone bosons $G^{0}$.}
\end{figure}
The total annihilation rate of (i) and (ii) are calculated as follows,
\begin{alignat}{1}
 &\Gamma=\langle\sigma_{1}v+\sigma_{2}v\rangle n_{\chi_{1}},\hspace{0.5cm} n_{\chi_{1}}=2\left(\frac{M_{\chi_{1}}T}{2\pi}\right)^{\frac{3}{2}}e^{-\frac{M_{\chi_{1}}}{T}}, \nonumber\\
 &\sigma_{1}v(\chi_{1}+\overline{\chi_{1}}\rightarrow G^{0}+G^{0})=\frac{M^{2}_{\chi_{1}}}{32\pi v_{2}^{4}}[1-\frac{7}{240}v^{4}+\cdots], \nonumber\\
 &\sigma_{2}v(\chi_{1}+\chi_{1}\rightarrow G^{0}+G^{0})=\frac{M^{2}_{\chi_{1}}}{128\pi v_{2}^{4}}[\frac{1}{(1-4y)^{2}}+\frac{1+4y}{4(1-4y)^{3}}v^{2}+\cdots], \nonumber\\
 &\langle\sigma_{1}v+\sigma_{2}v\rangle=a+b\,\langle v^{2}\rangle+c\,\langle v^{4}\rangle+\cdots\approx a+b\frac{6\,T}{M_{\chi_{1}}}\,,\nonumber\\
 &v=2\sqrt{1-\frac{4M^{2}_{\chi_{1}}}{s}}\,,\hspace{0.5cm} y=\frac{M^{2}_{\chi_{1}}}{M^{2}_{\phi_{2}^{0}}}\,,
\end{alignat}
where $v$ is a relative velocity of two annihilating particles. In view of Eq. (7) and Eq. (10), the thermal average on the annihilation cross-sections in Eq. (11) is exactly $\langle\sigma v\rangle\sim10^{-9}$ $\mathrm{GeV}^{-2}$, which is namely a weak interaction cross-section. This naturally reproduces the so-called ``WIMP Miracle" \cite{16}.

  As the universe temperature decreasing, the annihilation rate of $\chi_{1}$ becomes smaller than the Hubble expansion rate of the universe, then $\chi_{1}$ is decoupling. The freeze-out temperature is determined by
\begin{alignat}{1}
 &\Gamma(T_{f})=H(T_{f})=\frac{1.66\sqrt{g_{*}(T_{f})}\,T_{f}^{2}}{M_{Pl}}\,, \nonumber\\
\Longrightarrow &x=\frac{T_{f}}{M_{\chi_{1}}}\approx\left(17.6+ln\frac{M_{\chi_{1}}}{\sqrt{g_{*}(T_{f})x}}+ln\frac{\langle\sigma_{1}v+\sigma_{2}v\rangle}{10^{-10}\:\mathrm{GeV^{-2}}}\right)^{-1},
\end{alignat}
where $M_{Pl}=1.22\times10^{19}$ GeV, $g_{*}$ is the effective number of relativistic degrees of freedom. After $\chi_{1}$ is frozen out, its number in the comoving volume has no change any more. The current relic abundance of $\chi_{1}$ is calculated by the following equation \cite{16},
\ba
 \Omega_{\chi_{1}}h^{2}=\frac{0.85\times10^{-10}\:\mathrm{GeV}^{-2}}{\sqrt{g_{*}(T_{f})}\,x(a+3bx)}\approx 0.12,
\ea
where $a$ and $b$ are determined by Eq. (11). $0.12$ is the current abundance of the CDM \cite{17}. Obviously, $M_{\chi_{1}}$ and $v_{2}$ are jointly in charge of the final results of Eq. (12) and Eq. (13). Provided $M_{\chi_{1}}\sim30$ GeV and $v_{2}\sim300$ GeV, the solution of Eq. (12) is $x\sim\frac{1}{24}$, so $T_{f}\sim 1.3$ GeV. At this temperature the relativistic particles include $photon,gluon,G^{0},\nu^{0},e^{-},\mu^{-},u,d,s$, therefore we can figure out $g_{*}(T_{f})=62.75$. Finally, we can correctly reproduce $\Omega_{\chi_{1}}h^{2}\sim0.12$ by Eq. (13) . 

  The decoupling of the Goldstone boson $G^{0}$ is exactly at the same temperature
as one of the CDM $\chi_{1}$, obviously, it is much earlier than the neutrino decoupling and the photon one, thus the effective temperature of $G^{0}$ is lower than ones of the neutrino and the CMB photon. Therefore the current abundance of $G^{0}$ in the universe, $\Omega_{G^{0}}$, is smaller than the neutrino abundance $\Omega_{\nu}\approx1.7\times10^{-3}$ and the photon abundance $\Omega_{\gamma}\approx5\times10^{-5}$, refer to the review of cosmological parameters in \cite{1}. Since $G^{0}$ is massless and relativistic from its decoupling to the present day, now it should become a background radiation which is analogous to the CMB photon. However, we can not detect it through the ordinary methods because it does not interact with the SM matters.

  The effective potential between two CDM $\chi_{1}$ through the exchange of the Goldstone bosons is very complicated and unclear, but it should be a repulsive force because $\chi_{1}$ is a Majorana fermion, it is namely itself antiparticle. Therefore there are no any bound states of the CDM $\chi_{1}$. Two $\chi_{1}$ can happen elastic scattering via the $G^{0}$ mediation, moreover, the scattering cross-section is smaller than the weak interaction cross-section by one order of magnitude. When its reaction rate is smaller than the universe expansion rate, this elastic scattering will be frozen out and closed. The frozen-out temperature is determined by
\ba
\Gamma=\langle\sigma v\rangle n_{\chi_{1}}\approx\frac{M^{2}_{\chi_{1}}}{1024\pi v_{2}^{4}}\,\overline{v}\,n_{\chi_{1}}<H(T)
\ea
where $\overline{v}\approx\sqrt{\frac{2\,T}{\pi M_{\chi_{1}}}}$ is an average relative velocity. By use of the parameter values in Eq. (19), the frozen-out temperature is calculated as $\frac{T}{M_{\chi_{1}}}\approx\frac{1}{18}$. This temperature is slightly higher than the $\chi_{1}$ decoupling temperature $\frac{T_{f}}{M_{\chi_{1}}}\approx\frac{1}{24}$, obviously, the reason for this is that the annihilation cross-section in Eq. (14) is smaller than one in Eq. (11). Therefore, the elastic scattering between the CDM $\chi_{1}$ is actually frozen out before they are decoupling. Thereafter they are completely free particles except the gravitational influence. In conclusion, the model can simply account for the CDM, in particular, naturally explain the ``WIMP Miracle".

\vspace{0.6cm}
\noindent\textbf{IV. Leptogenesis}
\vspace{0.3cm}

  The model can also account for the baryon asymmetry through the leptogenesis at the scale of $v_{2}\sim v_{H}$. After $U(1)_{D}$ and $U(1)_{B-L}$ are broken one after another, the $D$ and $B-L$ quantum numbers of the heavy doublet scalar $\Phi$ become meaningless and vanishing. It can even mix with the SM Higgs $H$ since they have the same quantum numbers under the $G_{SM}$. In fact, the $B-L$ violation essentially arises from the last term in Eq. (6) when $\phi_{2}$ develops $\langle\phi_{2}\rangle\sim v_{2}$. Since all of $H,\phi_{1}^{0},\phi_{2}^{0}$ have no any $B-L$ numbers, the $B-L$ number of $\Phi$ also becomes meaningless and should be reassigned as zero. Thus the $B-L$ violation in the scalar sector is transferred to the Yukawa sector. $\Phi$ has two decay modes on the basis of the model couplings and Eq. (10), (i) the two-body decay $\Phi\rightarrow H+\phi_{2}^{0}$ and $\Phi\rightarrow H+G^{0}$, (ii) the three-body decay $\Phi\rightarrow l_{\alpha}+l_{\beta}+\overline{H}$, its tree and loop diagrams are shown as Fig. 2,
\begin{figure}
 \centering
 \includegraphics[totalheight=7cm]{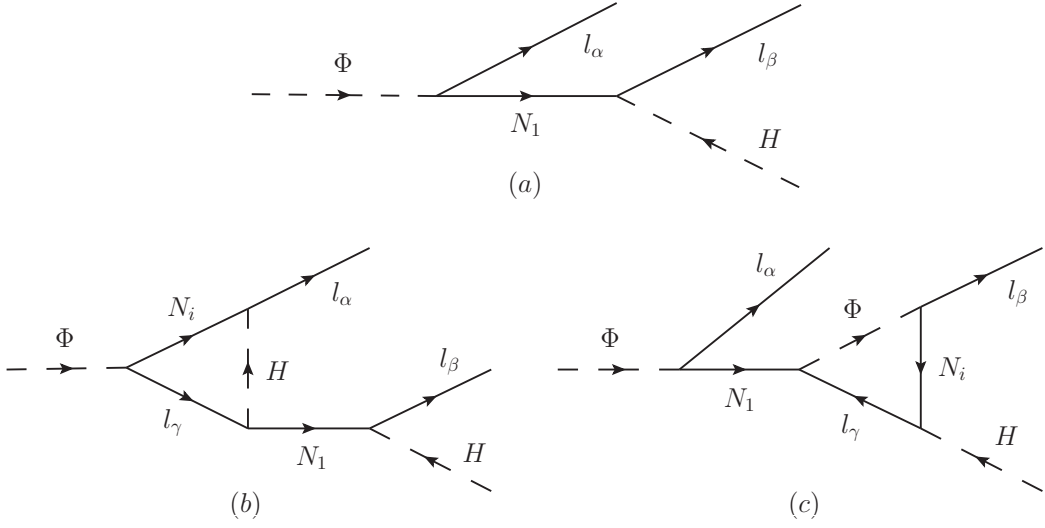}
 \caption{The tree and loop diagrams of the $B-L$ violating decay $\Phi\rightarrow l_{\alpha}+l_{\beta}+\overline{H}$, the decay is $CP$ asymmetric and out-of-equilibrium, which eventually leads to the matter-antimatter asymmetry.}
\end{figure}
explicitly, this process violates ``$-2$" unit of the $B-L$ number. Note that the three-body decay in Fig. 2 is mainly mediated by $N_{1}$, the decays via the $N_{2,3}$ mediation are greatly suppressed due to $M^{4}_{N_{2,3}}\gg M^{4}_{N_{1}}$, so they are ignored. Because the decay rate of (i) is much larger than one of (ii), the total decay width of $\Phi$ is approximately equal to the two-body decay width of (i).

 Because the couplings $Y_{1}$ and $Y_{2}$ contain the $CP$-violating sources, the decay rate of $\Phi\rightarrow l_{\alpha}+l_{\beta}+\overline{H}$ is different from one of its $CP$-conjugate process $\overline{\Phi}\rightarrow\overline{l_{\alpha}}+\overline{l_{\beta}}+H$ through the interference between the tree diagram and the loop one. The $CP$ asymmetry of the two decay rates is defined and calculated as follows,
\begin{alignat}{1}
 &\varepsilon=\frac{\Gamma^{+}-\Gamma^{-}}{\Gamma_{\Phi}}=\frac{(Y_{1}^{\dagger}Y_{1})_{11}M^{4}_{\Phi}\sum\limits_{i\neq1}M_{N_{i}}Im[(Y_{1}^{\dagger}Y_{2})_{1i}(Y_{2}^{\dagger}Y_{1})_{1i}]}{768\pi^{3}M^{3}_{N_{1}}(\lambda_{7}v_{1})^{2}}\,, \nonumber\\
 &\Gamma^{\pm}=\sum\limits_{\alpha,\beta}\Gamma(\frac{\Phi\rightarrow l_{\alpha}+l_{\beta}+\overline{H}}{\overline{\Phi}\rightarrow\overline{l_{\alpha}}+\overline{l_{\beta}}+H})=\Gamma_{tree}+\Gamma_{loop}^{\pm}\,,\hspace{0.5cm} \Gamma_{tree}=\frac{(Y_{1}^{\dagger}Y_{1})_{11}(Y_{2}^{\dagger}Y_{2})_{11}M^{3}_{\Phi}}{1536\pi^{3}M^{2}_{N_{1}}}\,, \nonumber\\
 &\Gamma_{\Phi}\approx\Gamma(\Phi\rightarrow H+\phi_{2}^{0})+\Gamma(\Phi\rightarrow H+G^{0})=\frac{(\lambda_{7}v_{1})^{2}}{8\pi M_{\Phi}}\,.
\end{alignat}
A careful calculation shows that the imaginary part of the loop integration factor of the (b) diagram is derived from the three-point function $Im[(C_{0}+C_{12})(M^{2}_{l_{\alpha}},s_{12},M^{2}_{\Phi},M^{2}_{N_{i}},M^{2}_{H},M^{2}_{l_{\gamma}})]$ $=\frac{2\pi i}{M^{2}_{\Phi}-s_{12}}$, where $s_{12}=(p_{\overline{H}}+p_{l_{\beta}})^{2}$, but the (c) diagram has actually no contribution to $\varepsilon$ because the imaginary part of its three-point function is vanishing. Provided $Y_{1}\sim Y_{2}\sim0.01$ and $\lambda_{7}\sim10^{-7}$ as the discussions in Section II, then we can roughly estimate $\varepsilon\sim 10^{-8}$ from Eq. (7) and Eq. (10), this is a reasonable and suitable value.

  In addition, the calculation shows that the decay rate $\Gamma^{\pm}$ in Eq. (15) is smaller than the universe expansion rate, namely
\ba
 \Gamma^{\pm}\approx\Gamma_{tree}<H(M_{\Phi})=\frac{1.66\sqrt{g_{*}}M^{2}_{\Phi}}{M_{Pl}}\,,
\ea
therefore the decay process of Fig. 2 is actually out-of-equilibrium. At the scale of $M_{\Phi}$ the relativistic states include $G^{0},\phi_{2}^{0},\chi_{i}$ besides all of the SM particles, so $g_{*}=114$ in Eq. (16).

  We have completely demonstrated that the decay process of Fig. 2 satisfies Sakharov's three conditions \cite{18}, as a consequence, a $B-L$ asymmetry can surely be generated at the scale of $v_{2}\sim v_{H}$. It is given by the following relation \cite{19},
\ba
 Y_{B-L}=\frac{n_{B-L}-\overline{n}_{B-L}}{s}=\kappa\frac{(-2)\varepsilon}{g_{*}}\,,
\ea
where $s$ is the entropy density and $\kappa$ is a dilution factor. If the decay is severe departure from thermal equilibrium, the dilution effect is very weak, then we can take $\kappa\approx1$. In addition, the dilution effect from $N_{1}\rightarrow\Phi+\overline{l_{\alpha}}$ is almost nothing because the $N_{1}$ number density is exponentially suppressed compared to the $\Phi$ one on account of $\frac{M_{N_{1}}}{M_{\Phi}}\sim10^{2}$.

  As long as the temperature is above $\sim100$ GeV \cite{20}, the electroweak sphaleron process can fully put into effect, thus it can convert a part of the $B-L$ asymmetry into the baryon asymmetry. This is expressed by the following relation,
\ba
 \eta_{B}=\frac{n_{B}-\overline{n}_{B}}{n_{\gamma}}=7.04\,c_{s}Y_{B-L}\approx 6.2\times10^{-10},
\ea
where $c_{s}=\frac{28}{79}$ is the sphaleron conversion coefficient in the model.  Note that only the SM particles participate in the sphaleron process at the scale of $v_{2}\sim v_{H}$, while $G^{0},\phi_{2}^{0},\chi_{i}$ are not involved in it since they are all singlets under the $G_{SM}$. $7.04$ is a ratio of the entropy density to the photon number density. $6.2\times10^{-10}$ is the current value of the baryon asymmetry \cite{21}. When the universe temperature falls below $\sim100$ GeV, the sphaleron process is closed and the baryon asymmetry is kept up to the present day. Finally, it should be stressed that the leptogenesis is realistically accomplished just above the electroweak scale in the model.

\vspace{0.6cm}
\noindent\textbf{V. Numerical Results and Discussions}

\vspace{0.3cm}
  We now show some concrete numerical results of the model. All of the SM parameters have been fixed by the current experimental data \cite{1}. Some new parameters in the model can be determined by a joint consideration of the tiny neutrino mass, the CDM abundance, and the baryon asymmetry. For the sake of simplicity, we only choose a set of typical values in the parameter space as follows,
\begin{alignat}{1}
 &v_{1}=5000\:\mathrm{TeV},\hspace{0.5cm} v_{2}=300\:\mathrm{GeV},\hspace{0.5cm} v_{H}=246\:\mathrm{GeV}, \nonumber\\
 &M_{\Phi}=5\:\mathrm{TeV},\hspace{0.5cm} M_{\phi_{2}^{0}}=200\:\mathrm{GeV},\hspace{0.5cm} M_{H^{0}}=125\:\mathrm{GeV}, \nonumber\\
 &M_{N_{3}}=\frac{v_{1}}{2}=2500\:\mathrm{TeV},\hspace{0.5cm} M_{N_{1}}=100\:\mathrm{TeV},\hspace{0.5cm} M_{\chi_{1}}=36\:\mathrm{GeV}, \nonumber\\
 &\lambda_{7}=10^{-7},\hspace{0.5cm} (Y_{1}^{\dagger}Y_{1})_{11}=(Y_{2}^{\dagger}Y_{2})_{11}=10^{-4}, \nonumber\\
 &(Y_{1}Y_{N}^{-1}Y_{2}^{T})_{33}=10^{-3},\hspace{0.5cm} Im[(Y_{1}^{\dagger}Y_{2})_{13}(Y_{2}^{\dagger}Y_{1})_{13}]=5.4\times10^{-7}.
\end{alignat}
The above values are completely in accordance with the model requirements discussed in Section II. Firstly $v_{2}$ and $M_{\chi_{1}}$ are determined by satisfying Eq. (12) and fitting the CDM abundance, secondly $v_{1}, M_{\Phi},M_{N_{1}}$ are determined by fitting the neutrino mass and the baryon asymmetry and satisfying Eq. (16), lastly the Yukawa couplings are chosen as reasonable and consistent values.

Now put Eq. (19) into the foregoing equations, we can correctly reproduce the desired results,
\ba
 m_{\nu_{3}}\approx 0.05\:\mathrm{eV},\hspace{0.5cm} \Omega_{\chi_{1}}h^{2}\approx 0.12\,,\hspace{0.5cm} \eta_{B}\approx 6.2\times10^{-10}.
\ea
These are in agreement with the current experimental data very well \cite{1}. Here we only give the upper bound of neutrino mass which is assumed as $m_{\nu_{3}}$. All of the experimental data of the neutrino masses and mixing angles can completely be fitted by choosing suitable texture of $Y_{1}Y_{N}^{-1}Y_{2}^{T}$. By use of Eq. (15), we can work out $\frac{\Gamma_{tree}}{H}\approx0.07$, this demonstrates that the decay of Fig. 2 not only satisfies the condition of Eq. (16), but also is severely out-of-equilibrium. Finally, it should be stressed that we do not make any fine-tuning in Eq. (19), only the two values of $M_{\chi_{1}}$ and $Im[(Y_{1}^{\dagger}Y_{2})_{13}(Y_{2}^{\dagger}Y_{1})_{13}]$ are accurately fixed in order to fit $\Omega_{\chi_{1}}h^{2}\approx0.12$ and $\eta_{B}\approx6.2\times10^{-10}$ respectively, while the rest of the parameters are roughly taken as the order of magnitudes.

  Fig. 3 shows the curve of the $B-L$ breaking scale $v_{2}$ versus the CDM mass $M_{\chi_{1}}$, which can correctly fit $\Omega_{\chi_{1}}h^{2}\approx0.12$.
\begin{figure}
 \centering
 \includegraphics[totalheight=8cm]{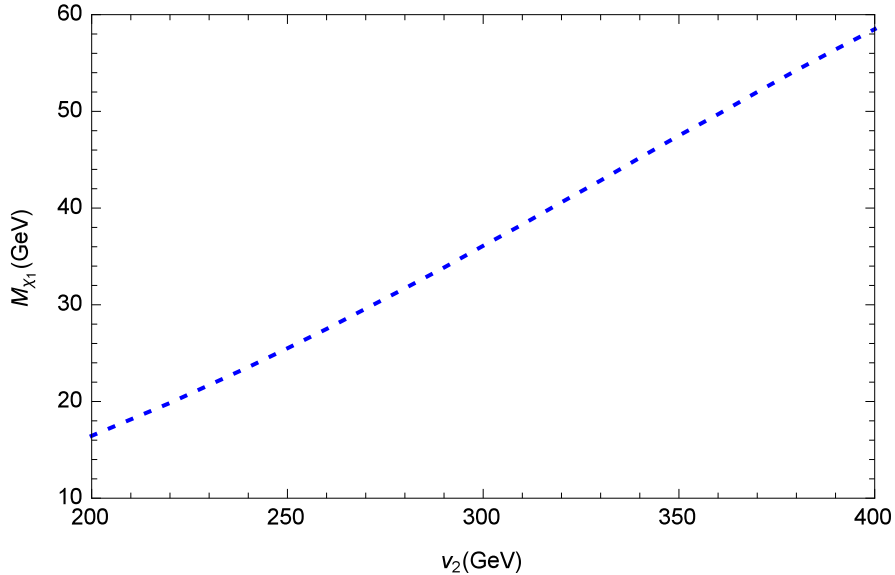}
 \caption{The curve of the $B-L$ breaking scale $v_{2}$ versus the CDM mass $M_{\chi_{1}}$, which can correctly fit $\Omega_{\chi_{1}}h^{2}\approx0.12$.}
\end{figure}
The value areas of $v_{2}$ and $M_{\chi_{1}}$ are reasonable and moderate. The curve clearly indicates that $v_{2}$ is just above the electroweak scale $v_{H}$, and the CDM $\chi_{1}$ mass is about several dozen GeVs. The experimental search for $\chi_{1}$ should therefore focus on this parameter area.

  In the end, we simply discuss the test of the model. Some new particles can be produced at the TeV-scale colliders. The relevant processes are listed below,
\begin{alignat}{1}
 &p+p \rightarrow\gamma+\gamma\rightarrow \Phi+\overline{\Phi},\hspace{0.5cm} e^{-}+e^{+}\:or\:p+\overline{p} \rightarrow\gamma\rightarrow \Phi+\overline{\Phi}, \nonumber\\
 &\Phi\rightarrow H+\phi_{2}^{0}\:or\:H+G^{0},\hspace{0.5cm} \Phi\rightarrow l_{\alpha}+l_{\beta}+\overline{H},\hspace{0.5cm} \phi_{2}^{0}\rightarrow \chi_{1}+\chi_{1}\:or\:G^{0}+G^{0}.
\end{alignat}
At the present LHC \cite{22}, we have a chance to search $\Phi$ and $\overline{\Phi}$ via two gamma photon fusion if the collider energy can reach their masses. Of course, a better way to produce $\Phi$ and $\overline{\Phi}$ is at $e^{-}+e^{+}$ or $p+\overline{p}$ colliders via the s-channel gamma photon mediation as long as the center-of-mass energy is enough high, for instance, the future colliders as CEPC and ILC have some potentials to achieve this goal \cite{23}. Only if $\Phi$ and $\overline{\Phi}$ are produced, then we can directly test the leptogenesis mechanism of the model by $\Phi\rightarrow l_{\alpha}+l_{\beta}+\overline{H}$ and $\overline{\Phi}\rightarrow \overline{l_{\alpha}}+\overline{l_{\beta}}+H$, on the other hand, this can indirectly shed light on the neutrino mass origin. In addition, we can probe $\phi_{2}^{0}$ and $G^{0}$ by $\Phi\rightarrow H+\phi_{2}^{0}$ and $\Phi\rightarrow H+G^{0}$. Finally, $\phi_{2}^{0}$ can decay into two CDM $\chi_{1}$ or $G^{0}$, by which we can measure the $\chi_{1}$ mass and find the Goldstone boson. All kinds of the final state signals are very clear in the decay chain of $\Phi$ and $\overline{\Phi}$.
  
  Of course, the CDM $\chi_{1}$ can be directly detected through scattering off nuclei at the underground detectors such as DAMA, XENON, etc. An indirect way is a search for the high-energy gamma photon and Goldstone boson in the cosmic rays, which are produced by the $\chi_{1}$ annihilation, shown as Fig. 4, but this detection is very difficult because its annihilation cross-section is too small. However, it will be very large challenges to actualize the above-mentioned experiments, this needs the researchers make a great deal of efforts. We will give an in-depth discussion on the model test in another paper.
\begin{figure}
 \centering
 \includegraphics[totalheight=5cm]{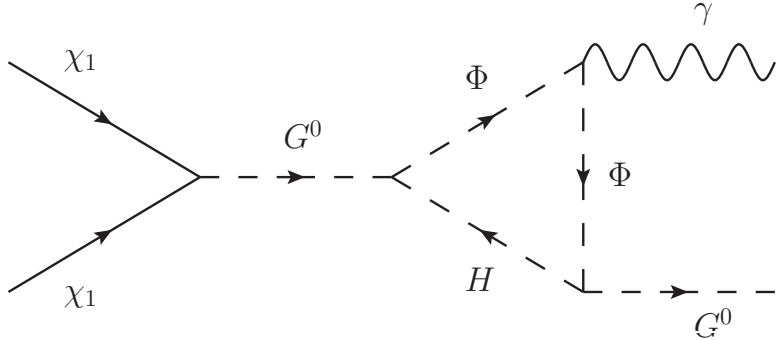}
 \caption{The indirect detection of the CDM $\chi_{1}$ through a search for the high-energy gamma photon and Goldstone boson in the cosmic rays.}
\end{figure}

\vspace{0.6cm}
\noindent\textbf{VI. Conclusions}

\vspace{0.3cm}
  In summary, I make an extension of the SM by the introduction of the dark sector with the local $U(1)_{D}$ symmetry. The particles in the dark sector have the non-vanishing $D$ numbers, while the SM particles are vanishing $D$ numbers. $U(1)_{D}$ is broken at the scale of thousands of TeVs, this gives rise to some particle masses in the dark sector. The global $B-L$ symmetry is violated just above the electroweak scale, this generates the CDM mass and leads to the ``WIMP Miracle", simultaneously, the leptogenesis is achieved by the decay of the dark doublet scalar $\Phi$ into two doublet leptons and one Higgs doublet anti-boson. The tiny neutrino mass is jointly caused by the heavy neutral Dirac fermion and the small vacuum expectation value of $\Phi$, the latter is induced from the very weak scalar coupling. In brief, the model is not complicated and its parameters are not many, but it can simultaneously account for the tiny neutrino mass, the CDM and the matter-antimatter asymmetry. Some interesting predications of the model, for example, the leptogenesis just above the electroweak scale, the CDM $\chi_{1}$ with the mass about several dozen GeVs, the background radiation of Goldstone bosons with the tiny abundance, are probably probed by the TeV collider experiments, the underground detectors, and the search in the cosmic rays. In short, these new physics beyond the SM are very attractive and worth researching in depth.

\vspace{0.6cm}
 \noindent\textbf{Acknowledgements}

\vspace{0.3cm}
  I would like to thank my wife for her large helps. This research is supported by the Fundamental Research Funds for the Central Universities Grant No. WK2030040054.

\end{document}